\documentclass[twocolumn,showpacs,preprintnumbers,amsmath,amssymb]{revtex4}

\usepackage{graphicx}
\usepackage{dcolumn}
\usepackage{bm}
\usepackage{axodraw4j}
\def\beq{\begin{equation}}
\def\eeq{\end{equation}}
\def\beeq{\begin{eqnarray}}
\def\eeeq{\end{eqnarray}}

\def\LQCD{\Lambda_{\mbox{\rm\scriptsize QCD}}}
\def\as{\alpha_{\mbox{\rm\scriptsize s}}}

\newcommand \Pomeron {I\!\!P}

\begin{document}

\title{Origins of parton correlations in nucleon and multi-parton collisions}

\author{B.\ Blok$^{1}$, Yu.\ Dokshitzer$^{2}$,   
L.\ Frankfurt$^{3}$
and M.\ Strikman$^{4}$
\\[2mm] \normalsize $^1$ Department of Physics, Technion -- Israel Institute of Technology,
Haifa, Israel
\\ \normalsize $^2$ LPTHE, University Paris--VI and CNRS, Paris, France\\
{\small On leave of absence: St.\ Petersburg Nuclear Physics Institute, Gatchina, Russia}\\
\normalsize $^3$ School of Physics and Astronomy,
\normalsize Tel Aviv University,
Tel Aviv, Israel
\\ \normalsize $^4$ Physics Department, Penn State University, University Park, PA, USA}

\begin{abstract}
We demonstrate that perturbative QCD leads to positive 3D parton--parton correlations inside nucleon explaining a factor two enhancement 
of the cross section of multi-parton interactions observed at Tevatron at  $x_i\ge 0.01$
as compared to the predictions of the independent parton approximation.  
We also find that though perturbative correlations decrease with $x$ decreasing, 
the nonperturbative mechanism kicks in and should generate correlation which, at $x$ below $10^{-3}$, is comparable 
in magnitude with the perturbative one for $x\sim 0.01$.
\end{abstract}

  \maketitle
\thispagestyle{empty}

{\em Multiple hard parton interactions}\/ (MPI) is an important element of the picture of strong interactions at high energies. 
At the LHC energies MPI with $p_\perp\sim$ few GeV occur in inelastic collisions with probability of the order of one.
The issue of MPI is attracting a lot of attention.

Building up on pioneering works of the early eighties \cite{TreleaniPaver,mufti}, 
theoretical studies were carried out in the last decade; for summary see 
proceedings of the MPI workshops~\cite{MPI-work,Bartalini:2011jp}.

Multi-parton interactions can serve as a probe for correlations between partons in the nucleon 
wave function and are crucial for determining the structure of the underlying event at LHC energies.  
MPI are currently modeled by Monte Carlo (MC) generators assuming the picture of independent partons.

A number of experimental studies were performed at the Tevatron \cite{Tevatron1,Tevatron2,Tevatron3}. 
New measurements are underway at the LHC, as MPI may constitute an important background for new physics searches.
The longstanding puzzling feature of the Tevatron data  is that the independent parton approximation which is strongly constrained by the HERA data on the hard diffractive vector meson production leads to at least a factor of two smaller cross section than reported experimentally \cite{Frankfurt}.
To describe the data, the MC models employ for the area of transverse localization of partons the value which is twice smaller than the one indicated by the HERA data.

Double hard parton scattering in hadron--hadron collisions contributes to production
of {\em four}\/ hadron jets with large transverse momenta $p_{i\perp}^2 \!\!\gg\!\! \LQCD^2$, of {\em two}\/ electroweak bosons
, or  ``mixed'' ensembles comprising three jets and $\gamma$, two jets and $W$, etc. 
For the sake of definiteness, we will refer to production of four final state jets.   

In \cite{BDFS1,BDFS2} we have developed a new formalism to address the problem of multi-parton interactions. 
QFT description of double hard parton collisions calls for introduction of a new  object --- 
the two-particle generalized parton distribution, $_2$GPD. 
Defined in the momentum space, it  characterizes non-perturbative two-parton correlations inside hadron \cite{BDFS1}.

The double hard interaction cross section (and, in particular, that of production of two dijets)
can be expressed through the generalized two-parton distributions $_2$GPD
\[
 D^{bc}_a(x_1,x_2,q_1^2,q_2^2,\Delta^2).
\] 
Here the index $a$ refers to the hadron, indices $b,c$ to the partons, $x_1$ and $x_2$ 
are the light-cone fractions of the parton momenta, and $q_1^2,q_2^2$ the corresponding hard scales.  
The two-dimensional vector $\vec\Delta$ is Fourier conjugate to the relative transverse distance 
between the partons $b$ and $c$ in the impact parameter plane.

The two partons can originate from the non-perturbative (NP) hadron wave function or, alternatively, emerge 
from perturbative (PT) splitting of a single parton taken from the hadron. 
In the first scenario one expects that the typical distance between partons is large, of the order of the hadron size $R$, 
so that the corresponding correlator in the momentum space is concentrated at small NP scale 
$\Delta^2\sim R^{-2}$ and falls fast at large momenta (exponentially or as a high power of $\Delta^2$). 
At the same time, PT production of the parton pair is concentrated at relatively small distances, 
so that the corresponding contribution to $_2$GPD is practically independent on $\Delta^2$ 
in a broad range up to the hard scale(s) characterizing the hard process under consideration.   

Separation of PT and NP contributions is a delicate issue. We suppose the existence of a separation scale $Q_0$ (of order of 1 GeV)
such that the $\Delta$-dependence of the correlation function differs substantially for the two mechanisms. 
Once this is done, it is reasonable to represent the $_2$GPD as a sum of two terms:
\beq\label{eq:2terms} \begin{split}
& D^{bc}_a  (x_1,x_2,q_1^2,q_2^2,\Delta^2) =  \cr
& {}_{[2]}D^{bc}_a(x_1,x_2,q_1^2,q_2^2,\Delta^2)
+ {}_{[1]}D^{bc}_a(x_1,x_2,q_1^2,q_2^2,\Delta^2) ,
\end{split}
\eeq
where subscripts ${}_{[2]}D$ and ${}_{[1]}D$ mark the first and the second mechanisms, correspondingly: 
two partons from the wave function
versus one parton that perturbatively splits into two.

As it has been explained in \cite{BDFS2}, 
a double hard interaction of two pairs of partons that both originate from {\em perturbative splitting}\/ 
of a single parton from each of the colliding hadrons, does not produce back-to-back dijets.  
In fact, such an eventuality corresponds to a one-loop correction to the usual $2\to4$ jet production process 
and should not be looked upon as multi-parton interaction.  

There are two sources of genuine multi-parton interactions: 
four-parton collisions described by the product of (PT-evolved) $_2$GPDs of NP origin, 
\begin{subequations}\label{eq:DD}
\beq
{}_{[2]}\!D_{a}(x_1,x_2;\vec\Delta)\,  { } _{[2]}\!D_{b}(x_3,x_4; -\vec\Delta) , 
\eeq
and three-parton collisions described by the combination 
\beq
_{[2]}\!D_{a} 
\, _{[1]}\!D_{b} 
  \>
+\, {}_{[1]}\!D_{a} 
\, _{[2]}\!D_{b}. 
\eeq
\end{subequations}
The latter corresponds to an interplay between the NP two-parton correlation in one hadron, 
and the two partons emerging from a PT parton splitting in another hadron ---  
$3\to4$ processes. 

Hard scattering of two pairs of partons in one hadron collision event is formally a rare process.
Compared with production of the same final state in usual two-parton interactions ({\em two-to-four}\/), 
contribution of double hard processes  ({\em four-to-four}\/) is small. 
It is suppressed as a power of the overall hardness of the process, 
$\LQCD^2/Q^2$, with $Q^2\sim p_\perp^2, M_{W\perp}^2$. 

However, in specific kinematical region the contribution of $4\to4$ processes may become comparable with that of $2\to4$. 
This happens in the ``back-to-back kinematics'' when four final state jets group into two pairs each of which has relatively small transverse momentum imbalance, $\delta_{12\perp}^2, \delta_{34\perp}^2 \ll Q^2$, with 
$\vec{\delta}_{ij\perp} \equiv \vec{p}_{i\perp} + \vec{p}_{j\perp}$. 

In \cite{BDFS2} QCD evolution equations for $_2$GPDs were derived in the leading collinear approximation, 
and differential distributions in $\delta_{13}$, $\delta_{24}$ were
presented in the form resembling the DDT formula for the $p_\perp$-distributions of massive lepton pairs (Drell--Yan process).

Differential distributions due to $4\to4$ and $3\to4$ processes exhibit {\em double collinear enhancement}\/: they peak at small {\em jet imbalances}, 
$\delta_{ik}^2\ll p_{i\perp}^2\simeq p_{k\perp}^2$, 
\begin{subequations}\label{eq:diffs}
\begin{eqnarray}\label{eq:diffs1}
\left. \frac{d\sigma}{dt_1dt_2\, d^2\delta_{13} d^2\delta_{24}} \right/  \frac{d\sigma}{dt_1dt_2} 
&\propto& \frac{\alpha_s^2}{\delta_{13}^2\,\delta_{24}^2},  \\
\label{eq:diffs2}
\left. \frac{d\sigma}{dt_1dt_2\, d^2\delta_{13} d^2\delta_{24}}  \right/  \frac{d\sigma}{dt_1dt_2} 
&\propto&  \frac{\alpha_s^2}{\delta'^2\,\delta^2}, 
\end{eqnarray}
\end{subequations}
where $\delta'^2 \equiv (\delta_{13}+\delta_{24})^2 \ll \delta^2 = \delta_{13}^2 \simeq \delta_{24}^2$.

\begin{figure}[h]
\begin{center}
\includegraphics[width=0.4\textwidth]{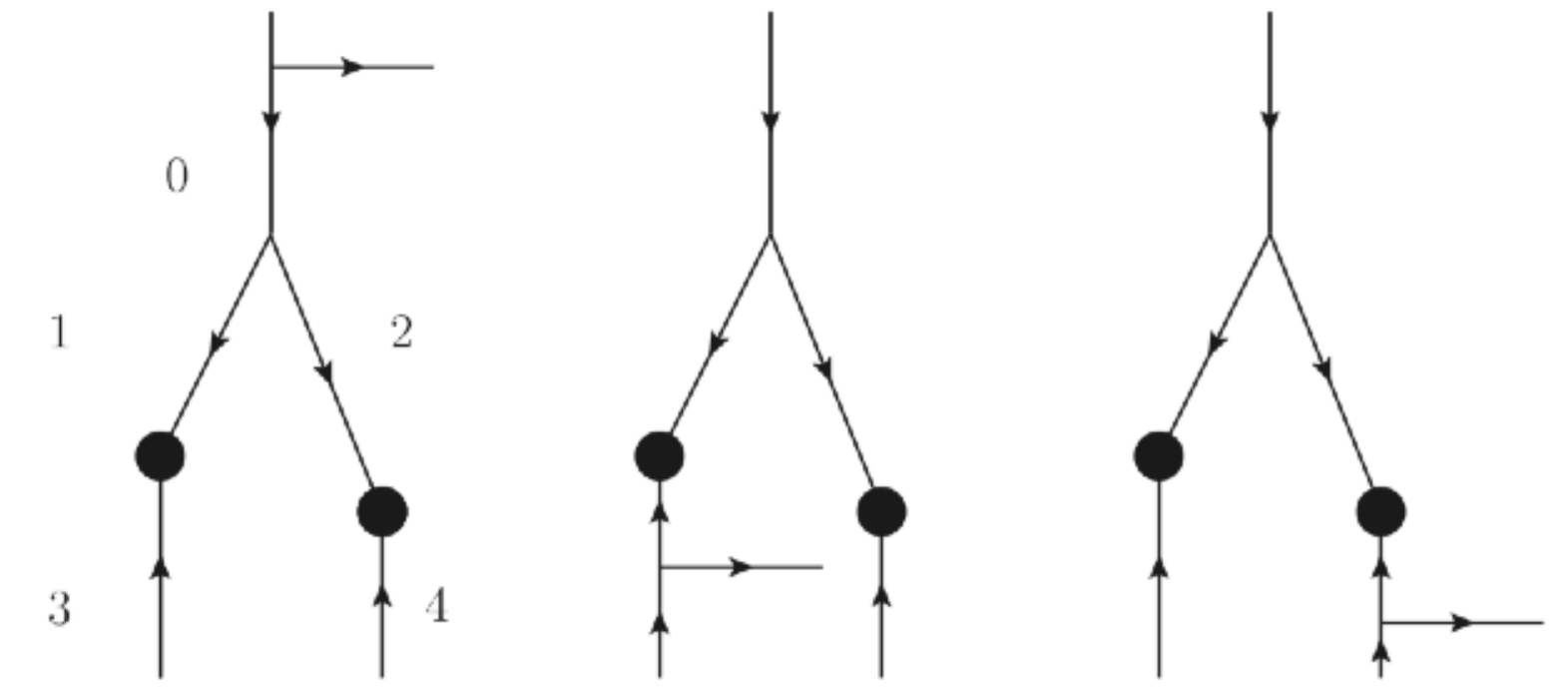}
\end{center}
\caption{\label{Fig8} Origin of $\delta'^2$ singularity in Eq.~\ref{eq:diffs2}}
\end{figure}

Structure of singularities displayed in Eq.~\ref{eq:diffs1} --- independent enhancements in two pair imbalances --- is typical for $4\to4$ processes.   
The $3\to4$ processes also contribute to Eq.~\ref{eq:diffs1} from the region of the transverse momentum scales of the splitting $\kappa^2\ll \max\{\delta_{13}^2,\delta_{24}^2\}$ (``internal splits'').
The situation is different when there is {\em no QCD emissions}\/ between the parton splitting $0\to 1+2$ and the two hard vertices as shown in Fig.~\ref{Fig8}. 
This ``end-point'' contribution is enhanced as Eq.~\ref{eq:diffs2}. 
Singularities in Eq.~\ref{eq:diffs} get smeared by double logarithmic Sudakov form factors of the partons involved, 
depending on the ratios of proper scales, see \cite{BDFS2}.

The four-jet production cross section due to double parton--parton scattering 
is conveniently represented as a product of cross sections of two independent hard collisions normalized 
by the {\em effective correlation area}\/ $S$ (known in the literature under a misleading name ``effective cross section''): 
\beq\label{eq:Dint}
   \frac{d\sigma^{(4)}}{dt_1 dt_2} = \frac{d\sigma(x_1,x_2)}{dt_1} \,\frac{d\sigma(x_3,x_4)}{dt_2} \times \frac1S.
\eeq
It is given by $\Delta$-integral of the product of $_2$GPDs Eq.~\ref{eq:DD}
\beq\label{eq:1S}
 \frac1S  = \!\int \frac{d^2\Delta}{(2\pi)^2} \,\left\{ {}_{[2]}\!D_{a}{}_{[2]}\!D_{b} + {}_{[1]}\!D_{a}{}_{[2]}\!D_{b} + {}_{[2]}\!D_{a}{}_{[1]}\!D_{b}\right\}.
\eeq
The PT $3\!\to\!4$ {\em end-point}\/ contribution to the total back-to-back cross section should be added to Eq.~\ref{eq:1S}, see Eqs.~(32) of \cite{BDFS2}.  
It {\em does not factorize}\/ into the product of two-parton distribution functions of colliding hadrons. 
In spite of being numerically small, it is worth trying to extract experimentally, as it manifests specific correlation between pair jet imbalances, Eq.~\ref{eq:diffs2}. 

\begin{figure}[b]  
\includegraphics[width=0.45\textwidth]{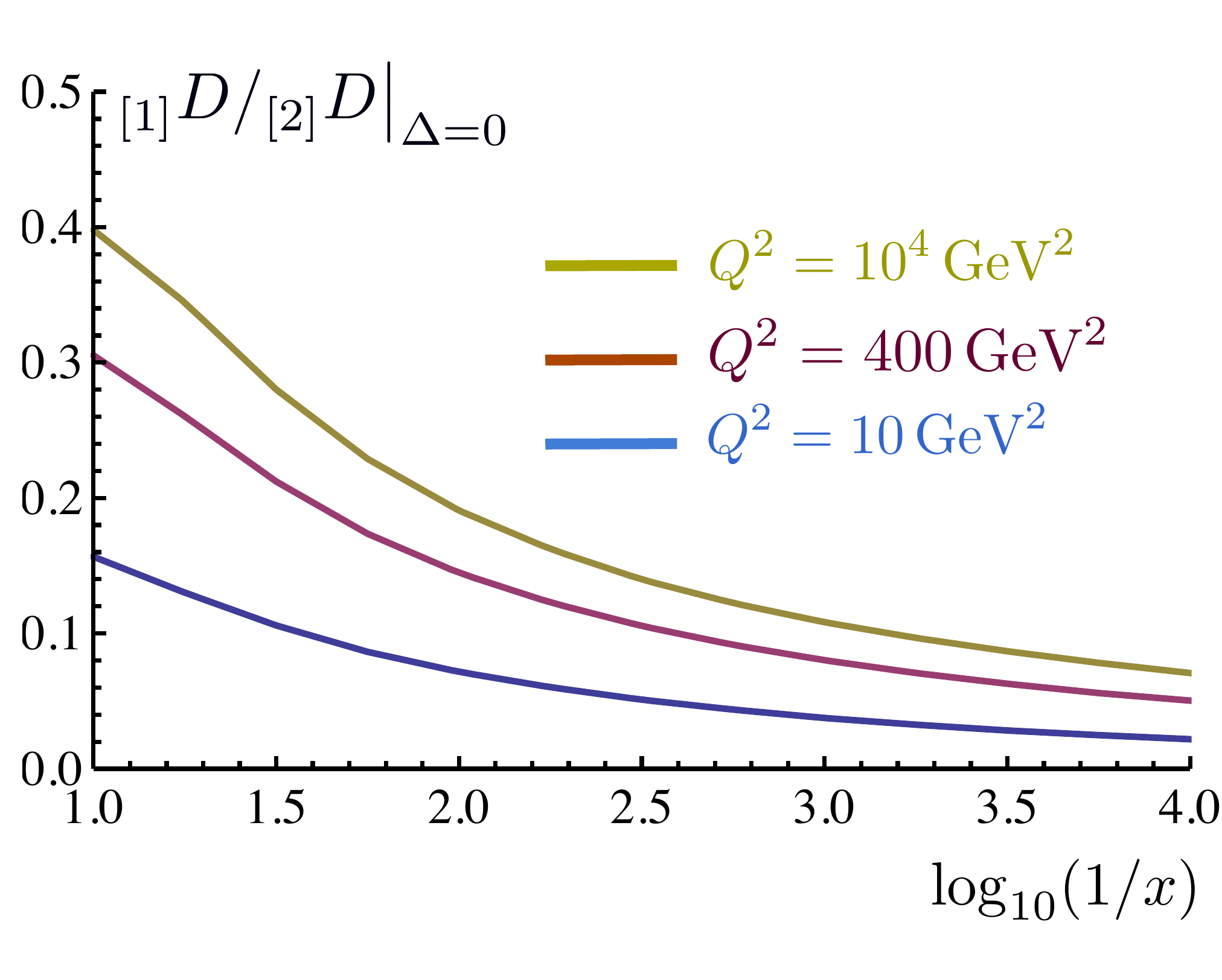}    
   \caption{The ratio of $3\!\to\!4$ to $4\!\to\!4$ PT-evolved contributions to $_2$GPD$(x_1\!=\!x_2,\Delta=0)$ for $Q_0^2\!=\! 1\,\mbox{GeV}^2$.}
    \label{2nd}
 \end{figure}

The NP two-parton distribution ${}_{[2]}\!D$ falls fast with $\Delta^2$ above $1\mbox{GeV}^2$, while ${}_{[1]}\!D$ depends on $\Delta$ only logarithmically. 
This enhances the contribution to the inverse effective correlation area of $3\to4$ processes (the sum of the second and third term in Eq.~\ref{eq:1S}) 
by about factor of {\em five}\/ as compared with the genuine $4\to4$ (the first term)~\cite{BDFS1}.    

In Fig.~\ref{2nd} the ratio ${}_{[1]}\!D_{p}/{}_{[2]}\!D_{p}$ at $\Delta\!=\!0$ is displayed for two gluons with $x_1\!\!=\! x_2$. 
It quantifies the strength of  {\em longitudinal}\/ correlation, since the point $\vec\Delta=0$ corresponds to integral over the transverse distance between partons.

The numerator was calculated by Eq.~18 of \cite{BDFS2} using the GRV parametrization of proton pdfs~\cite{GRV}. 
For the denominator we employed the evolution equation Eq.~16 of \cite{BDFS2} by taking for input 
the product of generalized one-parton distributions $D(x, q^2; t=-\Delta^2)$ according to the model of independent partons:
\begin{subequations}
\begin{eqnarray}
     {}_{[2]}\!D
     (x_1,x_2;q_1^2,q_2^2;\Delta) \!&=&\!\! D
     (x_1,q_1^2;\Delta^2) D
     (x_2,q_2^2;\Delta^2); \quad {} \\
     D
     (x,Q^2,\Delta^2) &=& G
     (x,Q^2)\cdot F_{g}(\Delta^2). 
\end{eqnarray}
\end{subequations}
Here $G$ is the standard gluon pdf and $F$ is the two-gluon form factor of the nucleon, see \cite{BDFS1}.
Fig.~\ref{2nd} indicates that the effective correlation area (``effective cross section'') does depend on the transverse momentum scale, 
an important feature which is not built in in the existing MC event generators.

The following simplified model allows one to get a qualitative estimate of the relative importance of the $3\!\to\!4$ contribution, 
as well as to understand its dependence on $x$ and the ratio of scales, $Q^2$ vs.\ $Q_0^2$.  
Imagine that at a low resolution scale, $Q_0$, the nucleon consisted of $n_q$ quarks and $n_g$ gluons (``valence partons'') with relatively large longitudinal momenta, 
so that triggered partons with $x_1,x_2\ll1$ resulted necessarily from PT evolution. 
In the lowest order, $\as\log (Q^2/Q_0^2)\equiv \xi$, the inclusive spectrum can be represented as
\[
    G \propto (n_qC_F + n_gN_c)\xi,
\]
where we suppressed $x$-dependence as irrelevant. 
If both gluons originate from the same ``valence'' parton, then 
\begin{subequations}
\beq\label{eq:1D}
{}_{[1]}\!D \propto  \frac12N_c\xi\cdot G +  (n_qC_F^2 + n_gC_FN_c)\xi^2,
\eeq
while independent sources give
\beq\label{eq:2D}
\begin{split}
{}_{[2]}\!D &\propto n_q(n_q\!-\!\!1)C_F^2 + 2n_qn_gC_FN_c +n_g(n_g\!-\!\!1)N_c^2 \cr
&= G^2 - (n_qC_F^2 + n_gC_FN_c)\xi^2 .
\end{split}
\eeq
\end{subequations}
Recall that the $\Delta$-dependence is different in Eq.~\ref{eq:1D} and  Eq.~\ref{eq:2D}. However, at $\Delta\!=\!0$
the second terms cancel in the sum and we get for the correlator
\beq\label{eq:corrmodel}
  \frac{D^{bc}(x_1,x_2;0)}{G^b(x_1) G^c(x_2)} -1 \>\simeq\> \frac{N_c}{2(n_qC_F+n_gN_c)}.
\eeq
The correlation is driven by the gluon cascade ---- the first term in Eq.~\ref{eq:1D}. It gets diluted when the number of independent ``valence sources'' 
at the scale $Q_0$ increases. This happens, obviously, when $x_i$ are taken smaller. On the other hand, for large $x_i\sim 0.1$ and increasing, 
the effective number of more energetic partons in the nucleon is about 2 and decreasing, so that the relative importance of the $3\to4$ processes grows.

\begin{figure}[h]  
  \vspace*{-0.1cm}
\includegraphics[width=0.45\textwidth]{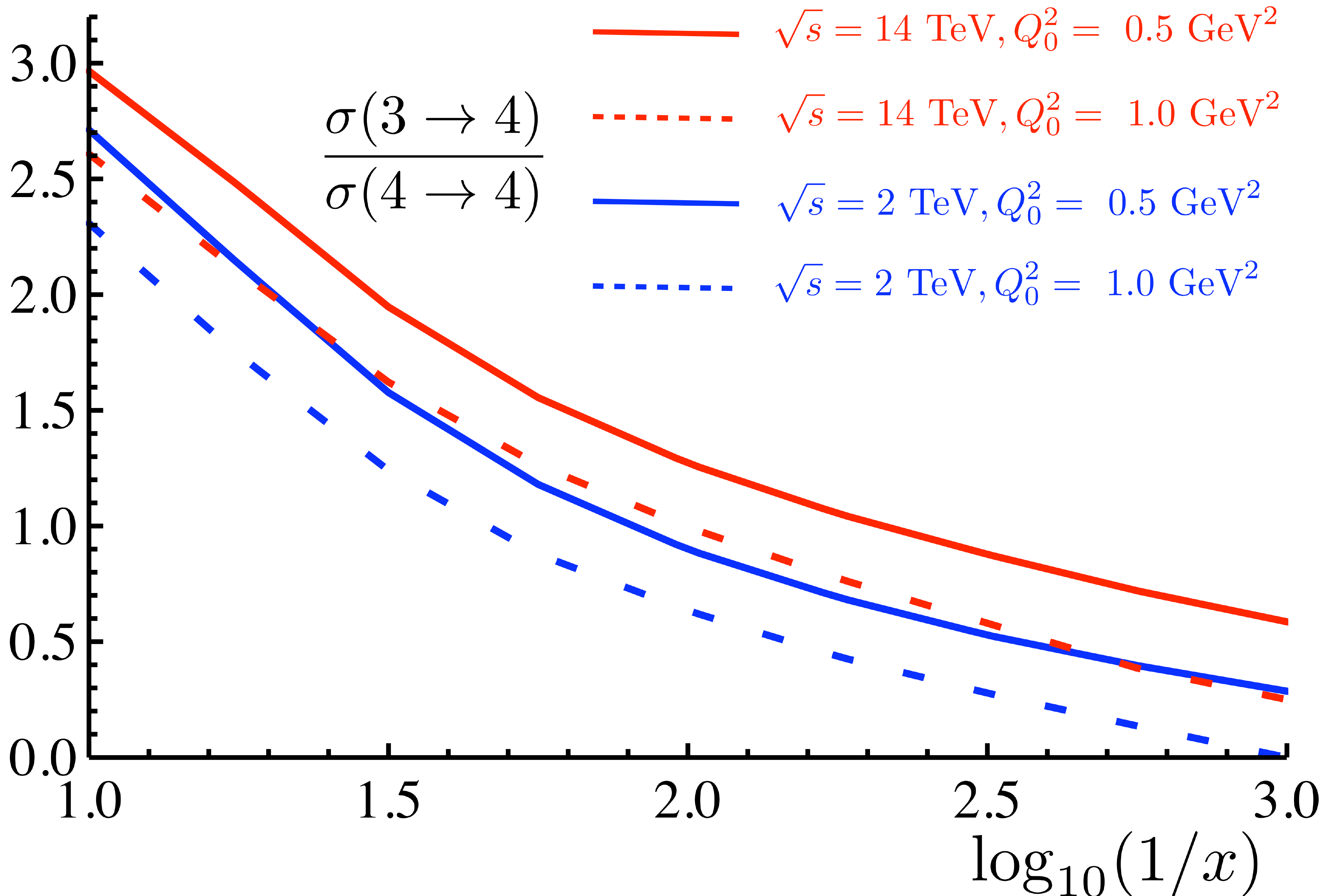}    
  \vspace*{-0.3cm}
   \caption{The ratio of $3\!\to\!4$ to $4\!\to\!4$ contributions to $4g\!\to\! 4\,\mbox{jets}$ cross section $(x_i=x)$ for Tevatron
   and LHC energies for two choices of the starting evolution scale $Q_0^2$; $Q^2=x^2s/4$.}
    \label{324}
 \end{figure}

It should be remembered that $3\to4$ mechanism contributes to the cross section $\simeq5$ times more than to the $_2$GPD in Eq.~\ref{eq:corrmodel}. 
Indeed, as Fig.~\ref{324} shows, at $x_i\simeq 10^{-2}$ (Tevatron)  the $3\!\to\!4$ contribution
enhances by about factor of 2 the four-jet production.

Thus, an account of $3\to4$ processes in combination with realistic one-parton GPDs explain the absolute magnitude of the cross section observed at the Tevatron
\cite{Tevatron1,Tevatron2,Tevatron3}. 
Fig.~\ref{324} is also consistent with the trend observed by D0~\cite{Tevatron3} of the increase of $1/S$ with 
$p_\perp$. 
A more informative confrontation of predictions with the data would require additional effort on both experimental and theoretical side.  

\medskip

\begin{figure}[h]  
\includegraphics[width=0.48\textwidth]{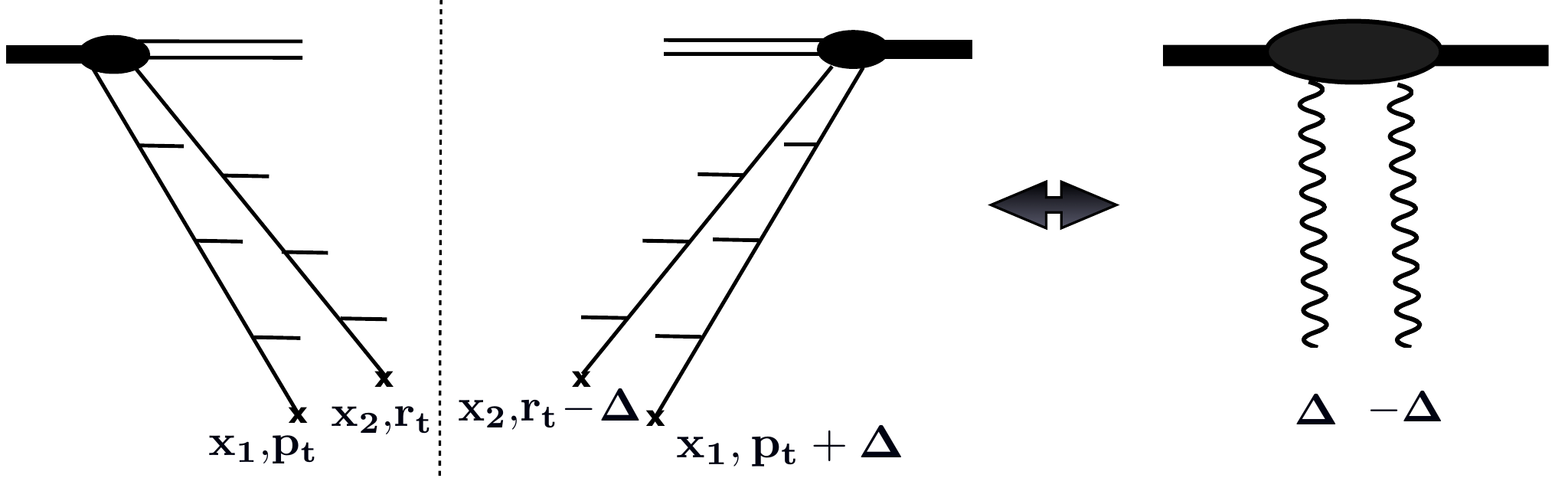} 
   \caption{ $_2$GPD as two-Pomeron exchange}
    \label{geom1}
 \end{figure}

A smooth matching between hard and soft QCD phenomena at  
$Q^2 \sim 1\, \mbox{GeV}^2$ allows one to assume that at such scale 
the single parton distributions at small $x$ below $10^{-3}$ are given by the soft Pomeron exchange.
In this picture the two soft partons originate from two independent ``multiperipheral ladders'' represented by {\em cut Pomerons}, see Fig.~\ref{geom1}.  

\begin{figure}[h]  
  \includegraphics[width=0.48\textwidth]{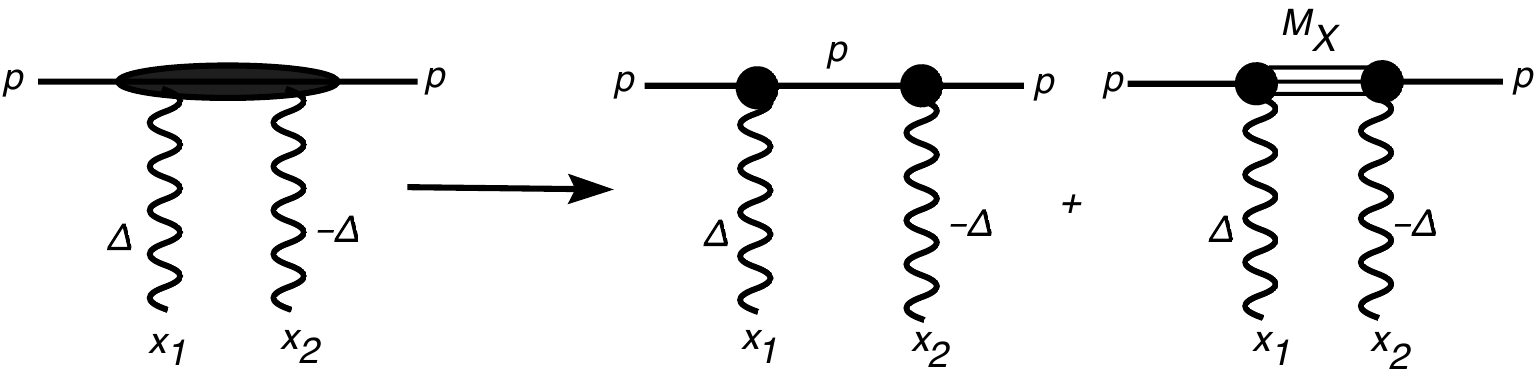} 
  \vspace*{-0.3cm}
   \caption{$2\Pomeron$ 
   contribution to $_2$GPD and Reggeon diagrams}
    \label{geom2}
 \end{figure}

 The soft Pomeron amplitude is practically pure imaginary.  
As a result, this amplitude equals that for the {\em diffractive cut}\/ of the two-Pomeron diagram of Fig.~\ref{geom2}. 
The  two contributions to the cut are the elastic and diffractive intermediate states. 
The elastic intermediate state obviously gives the uncorrelated contribution to $_2$GPD, while the inelastic diffractive cut encodes correlations.

It is convenient to introduce the ratio
\begin{equation}
R(x_1,x_2,-\Delta^2)={_{[2]}D(x_1,x_2;Q_0^2,Q_0^2;\Delta) \over  G(x_1,Q_0^2)\, G(x_2,Q_0^2) }.
\end{equation}
Then, Fig.~\ref{geom2} allows one to write  
\begin{equation}\label{elinel}
     R(x_1,x_2,t) = R_{\mbox{\scriptsize el}}(x_1,x_2,t)+ R_{\mbox{\scriptsize diff}}(x_1,x_2,t).
\end{equation}
The first term (uncorrelated partons) is given by the product of two single-parton GPDs,
\begin{equation}
R_{\mbox{\scriptsize el}}(x_1,x_2,t) = g_1(x_1, t)g_2(x_2, t), 
\end{equation}
with $g_i$ the ratio of a GPD to the corresponding pdf.  
The second term in Eq.~\ref{elinel} induces parton correlations, both transverse and longitudinal.   

By virtue of the QCD factorization theorem for hard diffraction \cite{harddiff}, the ratio $R_{\mbox{\scriptsize diff}}/R_{\mbox{\scriptsize el}}$ at $t\!=\!0$ 
that quantifies the strength of the longitudinal correlation is given directly by the ratio of the corresponding cross sections. The value
\beq
    \omega \equiv  {  \left. R_{\mbox{\scriptsize diff}}(x_1,x_2,0) \right/ R_{\mbox{\scriptsize el}}(x_1,x_2,0)} = 0.25\pm 0.05
\eeq
is extracted from the ratio of inelastic and elastic proton diffraction cross sections for electro-production of vector mesons studied at HERA \cite{Aaron:2009xp}. 
For $x<10^{-3}$ this parameter was found to depend only weakly on the incident energy ($x$). Moreover,
it stays roughly the same for light mesons and $J/\psi$ in a wide range of $Q^2$, thus confirming the hypothesis of a smooth transition between soft and hard regimes. 

Determination of the $t$-dependence of the ratio from the data is much more uncertain.  
Studies of various diffractive processes, both ``soft'' ($pp\to p +M_X$) and ``hard'' ($\gamma+p\to J/\psi+p$, $\gamma^*+p\to V+p$ with $V=\rho,\omega,\phi, J/\psi$)
indicate that the $t$-dependence of the differential cross section is dominated 
by the {\em elastic vertex}\/ $pp\Pomeron \propto \exp(B_{\mbox{\scriptsize el}}t)$ with $B_{\mbox{\scriptsize el}}=5\div 6\> \mbox{GeV}^2$ for $x<10^{-3}$. 
 
Using an exponential parameterization $\exp(B_{\mbox{\scriptsize inel}}t)$ for the square of the {\em inelastic vertex}\/ $pM_X\!\Pomeron$, 
the experimentally measured ratio of the slopes $B_{\mbox{\scriptsize inel}}/B_{\mbox{\scriptsize el}}  \simeq 0.28$ \cite{Aaron:2009xp} 
translates into the absolute value $B_{\mbox{\scriptsize inel}} = 1.4 \div 1.7\, \mbox{GeV}^2$. 
The fact that $t$-dependence of soft and hard inelastic processes is similar, goes in line, once again, with the logic of smooth matching of hard and soft regimes. 

The $\Delta$ integral of the first ($4\!\to\!4$) term in Eq.~\ref{eq:1S} gives the enhancement factor
\beq
\eta \equiv {(1/S)_{\mbox{\scriptsize corr}}\over (1/S)_{\mbox{\scriptsize uncorr}}}= 1 + 2 \omega {2B_{\mbox{\scriptsize el}} \over B_{\mbox{\scriptsize el}}+B_{\mbox{\scriptsize inel}}}
+\omega^2 {B_{\mbox{\scriptsize el}} \over B_{\mbox{\scriptsize inel}}}.
\eeq
\smallskip

The central value $\omega =0.25 $ yields $\eta= 2$.  
Hence, the MPI enhancement should persist even for $x<10^{-3}$ where we expect the PT correlations to diminish. 
\medskip

In conclusion, we have demonstrated that transverse and longitudinal correlations between partons in a nucleon generated by PT
and NP mechanisms play a critical r\^ole in explaining the observed absolute rate of MPI. 
Dedicated studies of the MPI at the LHC will provide a deeper understanding of the nucleon structure beyond single-parton distributions. 
It is pressing now for Monte Carlo models of $pp$ collisions to incorporate parton-parton correlations and employ realistic 
single-parton transverse space distributions as determined from hard exclusive processes at HERA. 

Acknowledgements:  This research was supported by the United States Department of Energy and the Binational Science Foundation.


\vfill 


\begin{thebibliography}{99}
\bibitem{TreleaniPaver}
  N.~Paver and D.~Treleani,
  Z.\ Phys.\  C {\bf 28}  187 (1985); \\
  N.~Paver and D.~Treleani,
  Nuovo Cim.\  A {\bf 70} (1982) 215.

\bibitem{mufti} M.~Mekhfi, Phys. Rev. D{\bf 32}, 2371 (1985).

\bibitem{MPI-work}  Proceedings of the 1st International Workshop
on Multiple Partonic Interactions at the LHC Verlug Deutsches Elektronen-Synchrotron, 2010.

\bibitem{Bartalini:2011jp} 
 P.~Bartalini, E.~L.~Berger, B.~Blok, G.~Calucci, R.~Corke, M.~Diehl, Y.~.Dokshitzer and L.~Fano {\it et al.},
 arXiv:1111.0469 [hep-ph].


%
%
%
%
%
%
%
%
%
%
%
%
%
%

\bibitem{Tevatron1}F.\ Abe {\it et al.}  [CDF Collaboration],
  Phys.\ Rev.\  D {\bf 56}, 3811 (1997).

\bibitem{Tevatron2}  V.M.\ Abazov {\it et al.}  [D0 Collaboration],
  Phys.\ Rev.\  D {\bf 81}, 052012 (2010).

\bibitem{Tevatron3}
V.M.\ Abazov {\it et al.}  [D0 Collaboration],
  Phys.\ Rev.\  D {\bf 83}, 052008 (2011).

	\bibitem{Frankfurt}
	  L.~Frankfurt, M.~Strikman and C.~Weiss,
	  Phys.\ Rev.\  D {\bf 69}, 114010 (2004); 
	  Ann.\ Rev.\ Nucl.\ Part.\ Sci.\  {\bf 55}, 403 (2005).




\bibitem{BDFS1}
  B.\ Blok, Yu.\ Dokshitzer, L.\ Frankfurt and M.\ Strikman,
  Phys.\ Rev.\  D {\bf 83}, 071501 (2011).

  \bibitem{BDFS2} B.~Blok, Yu.~Dokshitser, L.~Frankfurt and M.~Strikman,
  Eur.\ Phys.\ J.\ C {\bf72}, 1963  (2012). 



\bibitem{GRV}
  M.~Gluck, E.~Reya and A.~Vogt,
  Z.\ Phys.\  C {\bf 53}, 127 (1992).


\bibitem{harddiff}  
 J.~C.~Collins, L.~Frankfurt and M.~Strikman,
 Phys.\ Rev.\ D {\bf 56}, 2982 (1997).


\bibitem{Aaron:2009xp}  
 F.~D.~Aaron {\it et al.}  [H1 Collaboration],
 JHEP {\bf 1005}, 032 (2010).



\end{thebibliography}
\end{document}